\def\msol{M_{\odot}}

\documentclass[]{aa}
\usepackage{graphicx}
\begin{document}
\title{Energy release associated with a first-order phase transition in a rotating
neutron star core}

\author{J. L. Zdunik\inst{1}  \and
M. Bejger\inst{1,2}
 \and
P. Haensel\inst{1} \and
E. Gourgoulhon\inst{2} 
}
\institute{N. Copernicus Astronomical Center, Polish
           Academy of Sciences, Bartycka 18, PL-00-716 Warszawa, Poland
{\em jlz@camk.edu.pl}} \institute{N. Copernicus Astronomical Center,
Polish Academy of Sciences, Bartycka 18, PL-00-716 Warszawa, Poland
\and LUTH, UMR 8102 du CNRS, Observatoire de Paris, F-92195 Meudon Cedex, France\\
{\tt jlz@camk.edu.pl, bejger@camk.edu.pl, haensel@camk.edu.pl, Eric.Gourgoulhon@obspm.fr
 }}
\offprints{J.L. Zdunik}
\date{Received xxx Accepted xxx}
\abstract{}{We calculate energy release associated with a
first order phase transition at the center of a rotating
neutron star. This quantity is equal to the difference in
mass-energies between the initial normal phase configuration
and the final configuration containing a superdense matter
core, with total baryon number and angular momentum kept
constant. }{The calculations of the energy release are
based on precise numerical
2-D calculations, in which both the polytropic equations of
state (EOS) as well as realistic EOS of the normal phase are
used. Presented results are obtained for a broad range of metastability
of initial configuration and size of the new superdense
 phase  core in the final configuration.
When the equatorial radius of the dense core of the
superdense phase is  much smaller than the
stellar equatorial radius
analytical expressions for the energy release are obtained.}
{For a fixed ``overpressure'', $\delta\overline{P}$,
defined as the relative excess of central pressure of collapsing
metastable star over the pressure of equilibrium
first-order phase transition, the energy release
$\Delta E$ remarkably does
not depend on the stellar angular momentum and coincides with
that for nonrotating stars with the same $\delta\overline{P}$.
The energy release is  proportional
to $(\delta\overline{P})^{2.5}$ for small $\delta\overline{P}$,
when sufficiently precise brute force 2-D numerical calculations
are  out of question.
At higher $\delta\overline{P}$, results of 1-D calculations of
$\Delta E(\delta\overline{P})$ for non-rotating stars
are shown  to reproduce,  with very
high precision, the exact 2-D results for rotating stars.}{}

\keywords{dense matter -- equation of state -- stars: neutron -- stars: rotation}

\titlerunning{neutron-star core-quakes}
\maketitle
%
\section{Introduction}
\label{sect:introd}
 One of the intriguing predictions of some theories of dense matter
 in neutron-star cores  is a phase
transition into an ``exotic'' (i.e., not observed in
laboratory) state. Theoretical predictions include boson
condensation of pions and kaons, and deconfinement of quarks
(see, e.g., Glendenning 2000, Weber 1999).

As far as the structure and dynamics of neutron stars is
concerned, most important are the consequences of the
first-order phase transitions accompanied by discontinuities in the
thermodynamic potential densities. In the simplest case, one
considers states consisting of one pure phase. Because of high
degeneracy of matter constituents, effects of temperature are
neglected.  In thermodynamic equilibrium, phase transition
occurs then at a well defined pressure $P_0$, and is
accompanied by a density jump at the phase interface.

A first-order phase transition allows for a metastability of
the pure ``normal'' (lower density) phase at $P>P_0$.
Consequently, a metastable core could form during neutron-star
evolution in which central pressure increases, due to
accretion or spin-down. Then, nucleation of the exotic (higher
density) phase implies formation of a core of the exotic
phase and is accompanied by a core-quake and  energy release. A theory
which enables one to calculate the changes in stellar
parameters implied by a first-order phase transition in a
{\it non-rotating} neutron star was developed by Haensel et al.
1986 and Zdunik et al. 1987 (an earlier Newtonian theory was
presented by Schaeffer et al. 1983). The energy released in a
corequake was shown to depend strongly on the size of the
dense phase core, the leading term
being proportional to the fifth power of this core
radius.

In the present paper we calculate the energy release due to a
phase transition in a  {\it rotating} neutron star. Our theory is
based on the 2-D simulations and is much closer to reality
than the 1-D theory developed in  (Haensel et al. 1986, Zdunik
et al. 1987). In real world, evolutionary processes which lead
to the increase of the central density in neutron star
(accretion, slowing-down), as well as the collapse itself, all take
place in a rotating star. Of course, the 2-D calculations are
incomparably more difficult than the 1-D ones. However, as we
show in the present paper, when suitably parametrized,  the
energy released during a corequake essentially
depends only on the excess of the central pressure of
the metastable configuration over $P_0$, being to a good
approximation independent of the
angular momentum of collapsing star.

The paper is organized in the following way. In Sect.\ \ref{sect:EOS.theory}
 we introduce notations and describe general properties of the first-order
phase transitions in stellar core with particular emphasis on
the metastability and instability of neutron star cores.
Analytic considerations, concerning the response of a star to
a first-order phase transition at its center, and in particular,
the calculation , within the linear response approximation,
of the energy release associated with such a transition, are presented in
Sect.\ \ref{sect:E.lin.resp}.
Analytic models of the EOSs with first-order phase
transitions, allowing for very precise 2-D calculations, are
considered in Sect.\ \ref{sect:E.poly}, where we derive generic
properties of the energy release due to a first order phase transition
at the center of a rotating star. In Sect.\ \ref{sect:E.SLy}
we present our results obtained  for a realistic EOS of  normal
phase, and we confirm remarkable properties of the
energy-overpressure relation, obtained in the previous
section. In Sect.\ \ref{sect:num.estimates} we present
practical formulae suitable for the calculation of energy
release associated with a first order phase transition at the
center of a rotating neutron star. Finally,  Sect.\
\ref{sect:conclusions} contains discussion of our results
and several examples of application of our formula for the energy
release.
%
\section{EOS with a first  order phase transition }
\label{sect:EOS.theory}
 Let us consider a general case of a first-order phase transition
between the N ({\bf n}ormal)  and S ({\bf s}uperdense) phases
of dense matter. At densities under consideration, all constituents
of matter are strongly degenerate, and temperature dependence
of pressure and energy density can be neglected.
 At a given baryon density,  $n_{\rm b}$,  energy density of
 the N-phase of matter (including rest energies of particles which
 are matter constituents) is ${\cal
E}_{\rm N}(n_{\rm b})$ and pressure $P_{\rm N}(n_{\rm b})$.
The baryon chemical potential $=$ enthalpy per baryon  in the
N phase, is $\mu_{\rm N}=(P_{\rm N}+{\cal E}_{\rm N})/n_{\rm
b}$. Similarly, one can calculate thermodynamic quantities for
the S-phase.

\begin{figure}[h]
\centering
\resizebox{3.0in}{!}{\includegraphics[clip]{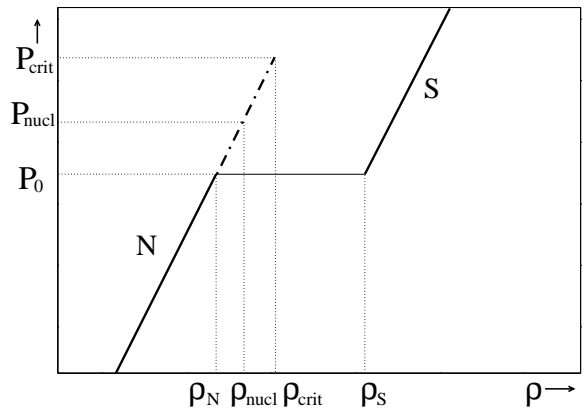}}
\vskip 2mm\caption{A schematic representation, in the $\rho-P$
plane, of an EOS with a first order phase
transition. Solid segments: stable N and S phase
(in thermodynamic equilibrium). Dash-dot segment:
metastable N phase. The S phase nucleates at $P_{\rm nucl}$, which
  depends on the temperature and the compression rate. At $P_{\rm crit}$
nucleation of the S phase is instantaneous, because the energy
barrier,  separating the N phase from the S phase,
vanishes. }
\label{fig:EOS1st}
\end{figure}
A proper thermodynamic variable, continuous  and monotonous  in
the stellar interior,   is the local pressure $P$. Equilibrium
state of the matter at a given $P$ is realized at minimum  of
enthalpy per baryon. For $P<P_0$, this minimum is realized by
the N phase, and for $P>P_0$ - by the S one. The value of
$P_0$ is obtained from the crossing condition $\mu_{\rm
N}(P)=\mu_{\rm S}(P)$, which yields  also the values of the
matter densities, $\rho_{\rm N}$ and $\rho_{\rm S}$, and the
corresponding baryon densities, $n_{\rm N}$ and $n_{\rm S}$,
at the N-S phase coexistence interface. These parameters are obtained
assuming thermodynamic equilibrium. A schematic plot of  EOS
of matter with a first order phase transition N-S,  in the
vicinity of the phase transition point, is plotted  in Fig.\
\ref{fig:EOS1st}.

Solid segment of the N-phase curve,  in Fig.\
\ref{fig:EOS1st},  corresponds to stable N-phase
state. For pressure above $P_0$, the N phase becomes
metastable with respect to the conversion into the S phase.
The S phase can appear through the nucleation  process -  a
spontaneous
formation of the S-phase droplets. However, an energy barrier
resulting from the surface tension at the N-S interface delays
the nucleation  for a time identified with a  lifetime of the
metastable state $\tau_{\rm nucl}$. The value of $\tau_{\rm nucl}$
decreases sharply with $P>P_0$, and drops to zero at some $P_{\rm
crit}$, where the energy barrier separating the S-state from
the N-state vanishes. For $P>P_{\rm crit}$ the N  phase is simply
unstable and converts with no delay into the S phase.

Consider a neutron star built of matter in the N phase.
Its central pressure $P_{\rm c}$ increases during spin-down or
accretion. A quasistatic compression of the N-phase core
moves matter into  a metastable state with $P>P_0$. A metastable core
of the N phase is bounded by a surface with $P=P_0$.  As soon
as the central compression timescale
metastable core $\tau_{\rm comp}=P_{\rm c}/\dot{P}_{\rm c}$ becomes equal
$\tau_{\rm nucl}$, droplets of the
 S phase appear at the star center. This happens at central pressure
 $P=P_{\rm nucl}<P_{\rm crit}$. The S-phase droplets introduce
 the pressure deficit,
 destabilize the metastable core and consequently - the  whole
 star. The S-phase core grows, and
 the process of the N$\longrightarrow$S phase transition proceeds until the
 final hydrostatic equilibrium state is reached. In the final
 state,
 a central core of the S-phase is bounded by an  N-S coexistence
 surface,  of constant pressure $P_0$,  on which
 matter density suffers a jump from $\rho_{\rm _N}$ on the
 outer N-phase side to $\rho_{\rm _S}$ on the inner S-phase side.

 In the present paper we restrict ourselves to the case of the
 density jump satisfying
 $\rho_{\rm S}<{3\over 2}(\rho_{\rm  _N}+P_0/c^2)$.
 Therefore, stellar configurations with a core of
 the S phase of arbitrarily small radius are stable (Seidov 1971,
 see also Kaempfer 1981 and Zdunik et al. 1987).
 The case of a strong first order phase transition with
 $\rho_{\rm S}>{3\over 2}(\rho_{\rm  _N}+P_0/c^2)$,  when
 configurations with a small S phase core are unstable
 and collapse into those with a large  S phase core,
 will be presented in a separate paper.

A metastable stellar state of the N phase can be described by central  {\it
overcompression} - fractional excess of density relative  to $\rho_{\rm
_N}$,
\begin{equation}
\label{rhocrit}
\delta\overline{\rho}\equiv {{\rho_{\rm
c}-\rho_{\rm _N}}\over \rho_{\rm _N}}~.
\end{equation}
Equivalently, we define a dimensionless central  {\it
overpressure},

\begin{equation}
\label{pcrit} \delta\overline{P} \equiv  {{P_{\rm
c}-P_0}\over P_0}\;.
\end{equation}

Let us now consider the timescales and their interplay.
First,  there is a  {\it microscopic} nucleation timescale
$\tau_{\rm nucl}(\delta\overline{P})$. The time dependence
$\delta\overline{P}(t)$ results from a {\it global}
evolution of neutron star and depends on the astrophysical
scenario of central compression; it will be characterized
by  $\tau_{\rm comp}=P_{0}/\dot{P}_{\rm c}=
1/\dot{\delta\overline{P}}$.
The time needed to compress matter in the center of the star
from $P_0$ to  $P_{\rm c}$ is, in linear approximation, equal
to  $\tau_{\rm comp} \cdot\delta\overline{P}$.
During this time the system remains in
a metastable state.
 In contrast, the functional
dependence $\tau_{\rm nucl}(\delta\overline{P})$ is given
by the  {\it local} dense matter microphysics
and is a very sensitive function
of $P$, being very large for $P$
close to $P_0$ and dropping down abruptly
above  some pressure
(see for example Iida \& Sato \cite{IidaSato1997}).
As a result,  the condition for the
time $t=t_{\rm quake}$ at which the metastable stellar
configuration collapses can be estimated by solving the equation:
\begin{equation}
\tau_{\rm nucl}\left[\delta\overline{P}(t)\right]
=q\cdot\tau_{\rm comp} \cdot\delta\overline{P}~.
\label{eq:t.quake}
\end{equation}
where a small dimensionless
coefficient $q$ reflects the very steep character of the function
$\tau_{\rm nucl}(\delta{\overline P})$
and is of the order $10^{-2}-10^{-3}$ (details will be
presented in our forthcoming paper).

\section{Calculation of the energy release}
\label{sect:E.num}
We restrict ourselves to axially symmetric, rigidly rotating neutron stars
in hydrostatic equilibrium. In what follows by ``radius'' we mean
a circumferential radius in an equatorial plane.

\begin{figure}[h]
\centering\resizebox{3.in}{!}{\includegraphics[clip]{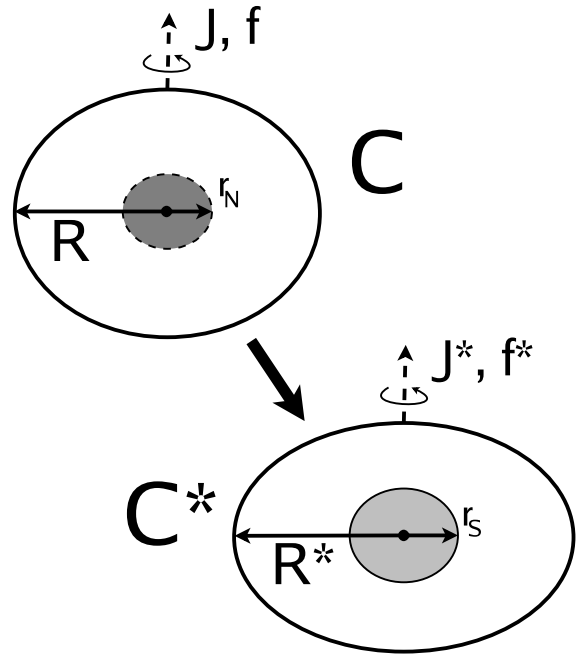}}
\caption{Transition
from a one-phase configuration ${\cal C}$
 with a meta-stable core of radius $r_{\rm _N}$ to a
two-phase configuration ${\cal C}^\star$
 with  S-phase core with a radius $r_{\rm
_S}$. By ``radius'' we mean an equatorial circumferential
radius. These two configurations have the same baryon number $A=A^\star$
and total angular momentum $J=J^\star$.}
\label{fig:CCstar}
\end{figure}

We assume that at a central pressure $P_{\rm c}=P_{\rm nucl}$
the nucleation of the S phase in an overcompressed core, of
radius $r_{\rm _N}$, of configuration ${\cal C}$, initiates the
phase transition and formation of S-phase core of radius
$r_{\rm _S}$ in a new configuration ${\cal C}^*$, as presented
on Fig.\ \ref{fig:CCstar}. Transition to an  S phase at the
core boundary, occurring at $r_{\rm _S}$,  is associated with a density
jump characterized by $\lambda_\rho=\rho_{\rm _S}/\rho_{\rm _N}$.
Because we are
interested in relatively small S-phase cores, we will
approximate its EOS by a polytrope with the exponent
$\gamma_{\rm _S}$, equal to the adiabatic index
of the S phase at $\rho=\rho_{\rm _S}$.
 Of course, in reality the adiabatic index of
the S phase does depend on the density. However, for  small  cores the
polytropic EOS is an excellent approximation, and in the limit when
only  leading terms in the $r_{\rm _S}$-expansion are kept this
approximation becomes
exact (Bejger et al. 2005).

Having a pair of EOSs, one with and one without a
softening by phase transition, the next step is to compare the hydrostatic equilibria
of neutron stars corresponding to each of these  EOSs. The
models which we calculate  are rigidly rotating, axisymmetric
solutions of Einstein's equations.
The numerical computations have been performed
by means of a code built on the Lorene library
({\tt http://www.lorene.obspm.fr}), with the accuracy of $10^{-6}$ or better 
measured using general relativistic virial theorems.
The neutron-star models can be labeled by the
central density $\rho_{\rm c}$ and rotational frequency $f=\Omega/2\pi$.
These parameters are natural from the point of view of numerical calculations
($\rho_{\rm c}$ and $f$ are input parameters in the numerical code).
But we can imagine another
parametrization, more useful for other purposes.
For example, to study the stability
of rotating stars,  the better choice is central density, $\rho_{\rm c}$,
 and total angular momentum of the star,  $J$.

\begin{figure}[h]
\centering
\resizebox{3.25in}{!}{\includegraphics[clip]{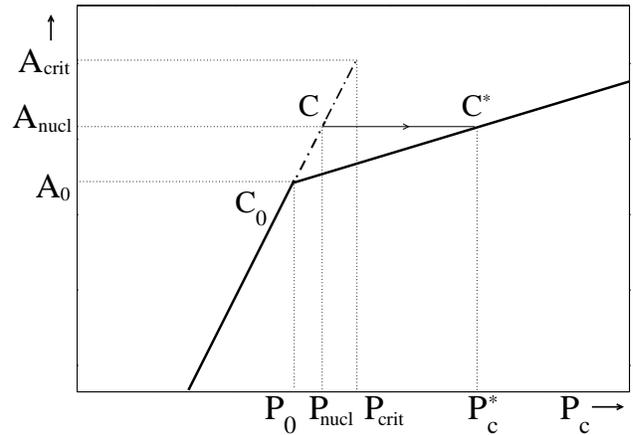}}
\caption{Total baryon number $A$ of hydrostatic stellar
configuration,
versus central pressure  $P_{\rm c}$, at fixed stellar
angular momentum $J$, for the EOS, depicted in
Fig.\ \ref{fig:EOS1st}. Solid line
denotes stable , dash-dot  line - the states which a
meta-stable with respect to the N$\longrightarrow$S transition.
For a central pressure $P_{\rm nucl}$ the
S-phase nucleates in the super-compressed core of
configuration ${\cal C}$, and this results in a transition ${\cal
C}\longrightarrow {\cal C}^*$ into a stable configuration with
a S-phase core and central pressure $P^*_{\rm c}$. Both
configurations  ${\cal C}$ and ${\cal C}^*$ have the same
baryon number $A$.} \label{fig:CCstarPc}
\end{figure}

Below $\rho_{\rm _N}$, the EOS for two cases (with and without phase transition)
is the same.
For the problem of non-rotating stars, considered
 two decades ago
papers (Haensel et al. 1986, Zdunik et al. 1987),
 the configuration with $\rho_{\rm c}=\rho_{\rm
_N}$ was denoted by ${\cal C}_0$, and treated as a
``reference configuration''.
Configurations ${\cal C}_0$, ${\cal C}$, and ${\cal C}^\star$
are  depicted in  the $A-P_{\rm c}$ plane  in Fig.\
\ref{fig:CCstarPc}. The radius, gravitational mass,
and the total baryon number of ${\cal C}_0$
will be denoted by $R_0$, $M_0$, and $A_0$.
For rotating stars we do not have
one "reference configuration" but a set of "reference
configurations" which depend on rotation rate
$\lbrace{\cal C}_0 (f)\rbrace$
or total angular momentum $\lbrace{\cal C}_0 (J)\rbrace$
of initial metastable configuration. As it has been
discussed in Sect.\ 2,  there exists also a  set of
configurations resulting from the existence of
metastability of the matter in N phase. The maximum central
pressure which can be reached in the N-phase star is defined
by the value $P_{\rm c}=P_{\rm nucl}$. For $P_{\rm c}>P_{\rm
nucl}$,  the phase transition in the center of the star takes
place on a timescale much shorter than time of the stellar
evolution (accretion or spin down). Thus the "critical
line" defined by the condition  $P_{\rm c}=P_{\rm nucl}$
corresponds to the configurations for which nucleation in the
center triggers  a collapse of the whole star and a corequake.
>From the astrophysical point of
view,  the phase transition in the center followed by a
corequake takes place at a point of a  "critical line".
Consequently,  "normalization" of results with respect to
parameters of "reference configuration" is more
complicated than in the case of non-rotating stars. There are many
possibilities to define "reference configuration" for a star
with metastable core ($P_{\rm c} > P_{\rm _N}$), rotating with
frequency $f$ and possessing a  total angular momentum $J$ and total
baryon number $A$. Such a  "reference configuration" should be defined as the
configuration with the central pressure equal to $P_{\rm 0}$,
as shown in Fig.\ \ref{fig:CCstarPc}.
However, we can choose either a reference configuration  star with the
same $J$ or the same $f$
as that of collapsing metastable configuration ${\cal C}$.  In fact,
 to reach the metastable configuration
with $P_{\rm c} > P_{0}$,  we have to cross the "reference
line" at the point defined by the evolutionary process leading to the instability.
Properties of ``critical line'' and ``reference line'' in
the space of rotating configurations of slowing down or
accreting neutron stars will be analyzed in more detail in our
forthcoming  paper.

We additionally assume,  that transition of the star from a  one-phase configuration to the
configuration with a small dense core built of S phase, takes place at fixed baryon
number $A$ (no matter ejection)  and fixed total angular momentum of the star $J$
(radiation loss of $J$ neglected).

\begin{figure}[h]
\centering \resizebox{\hsize}{!}{\includegraphics[angle=0]{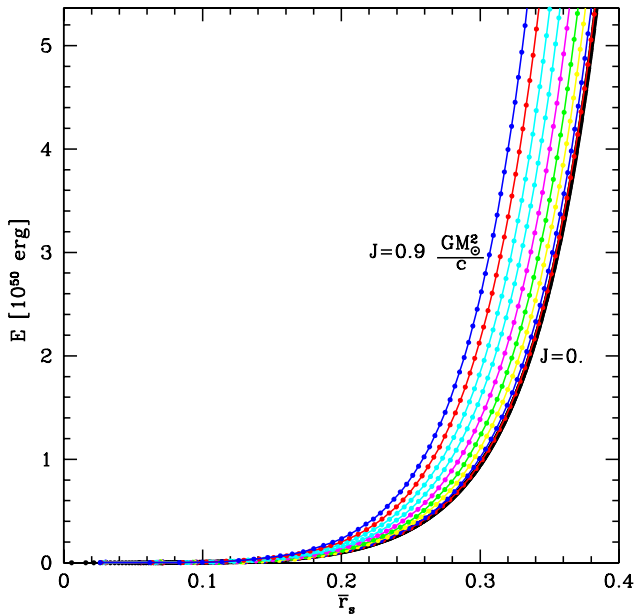}}
\caption{The energy release due to the corequake  of
rotating neutron star as a function of the dimensionless
equatorial radius of the new phase core,
$\overline{r}_{\rm _S}$. Different  curves
 correspond to the different values of total angular momentum of
rotating star, fixed along each curve, $J=(0, 0.1,~\ldots,~0.9)\times
GM^2_\odot/c$, from the bottom curve to the top curve.
 }
\label{fig:E.rs.poly}
\end{figure}

The energy release during transition
${\cal C}(A,f)\longrightarrow {\cal
C}^\star(A,f^\star)$ is calculated from the change of the stellar mass-energy
during this process,
\begin{equation}
\Delta E=\left[M({\cal  C })-M({\cal C}^\star)\right]_{A,J}c^2
\label{eq:DeltaE}
\end{equation}

\subsection{Energy release for polytropic EOSs}
\label{sect:E.poly}
%
In the present section we intend to simplify the analysis, removing
problems connected with numerical precision.  To this aim, we will use the
polytropic EOSs for the N and S phases. The polytropic EOSs
not only guarantee  high precision
of numerical calculation, but also open a possibility
of the exploration of wide region of the parameter space.
Description of the polytropic EOSs
and their application to
relativistic stars with phase transitions was
presented in detail in our
previous publications in this series (Bejger et al. 2005,
Zdunik et al. 2006).

\begin{figure}[h]
\centering \resizebox{\hsize}{!}{\includegraphics[angle=0]{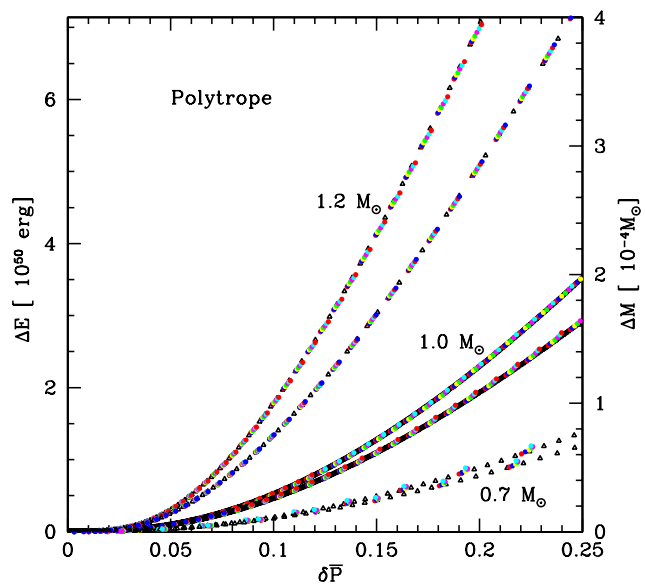}}
\caption{(Color online) The energy release due to the mini-collapse of
rotating neutron star as a function of the overpressure $\delta\overline{P}$
of the N phase of the matter
in the center of the star,  for the polytropic EOS ($\gamma=2$) with
first order phase transition located at three different points (pressures $P_0$)
for which the mass of the reference nonrotating configuration is equal
to $0.7,~1.0,~1.2\msol$. For each value of $P_0$ two models of S phase EOS are
considered (one of examples of the EOS depicted in
Fig.\ \ref{fig:EOS.SLy.poly}). They correspond to the S phase
described  by the  polytropes with
$\gamma_{\rm _S}= 2.5$ (upper curve) and
 $\gamma_{\rm _S}= 3$ (lower curve).
The points of different color correspond to the different values of total angular momentum of
rotating star, $J=(0, 0.1,~\ldots,0.9)\times GM^2_\odot/c$.
For a given EOS,  results for all rotating configurations can be very well
approximated by a single curve, independent of $J$. }
\label{fig:E.pc.poly}
\end{figure}

In Fig.\ \ref{fig:E.rs.poly} we presented the energy release
as a function of  $\overline{r}_{\rm _S}=r_{\rm _S}/R({\cal C})$,
for several values of the angular
momentum of the metastable configurations ${\cal C}(J)$. The reference
configuration is ${\cal C}$ - the initial configuration of the corequake
with metastable core in the center and central pressure $P_{\rm c}=P_{\rm nucl}$.
As we see in Fig.\ \ref{fig:E.rs.poly}, the energy release
corresponding to a given value of $\overline{r}_{\rm _S}$
depends rather strongly on the
rotation rate (here presented for the fixed values
of total angular momentum).

In Fig.\ \ref{fig:E.pc.poly} we presented the energy release
as a function of the overpressure of the metastable N phase in the center of the
metastable star ${\cal C}(A,J)$,
$\delta\overline{P} = P_{\rm nucl}/P_0-1$, for several values
of $J$.
As we already stressed, the value of $P_{\rm nucl}$ (or $\delta\overline{P}$)
 can be determined from microscopic considerations,
 combined with physical conditions prevailing
at the star center as well as
with their time evolution rate. Having determined
$P_{\rm nucl}$, we can  determine the energy release, $\Delta E$,  due to the
corequake ${\cal C}(A,J)\longrightarrow {\cal C}^\star(A,J)$,
where the metastable one-phase configuration, and the final two-phase
configuration, have the same values of the baryon number $A$
and total angular momentum $J$, which are conserved during the
transition.

As we see in Fig.\ \ref{fig:E.pc.poly}, the energy release
in a  (mini)collapse of rotating star is independent of
rotation rate of collapsing configuration, and  depends
exclusively on the degree of metastability of the N phase at the stellar
center (departure of matter from chemical equilibrium), measured by
the overpressure $\delta\overline{P}$. In particular,
 to calculate the energy release associated with a corequake of a
 rotating neutron star, it is sufficient to know the value of
 $\Delta E$  for a non-rotating star of the same central overpressure,
  which can be deduced from expressions derived in \cite{ZHS}.

\begin{figure}[h]
\centering
\resizebox{\hsize}{!}{\includegraphics[]{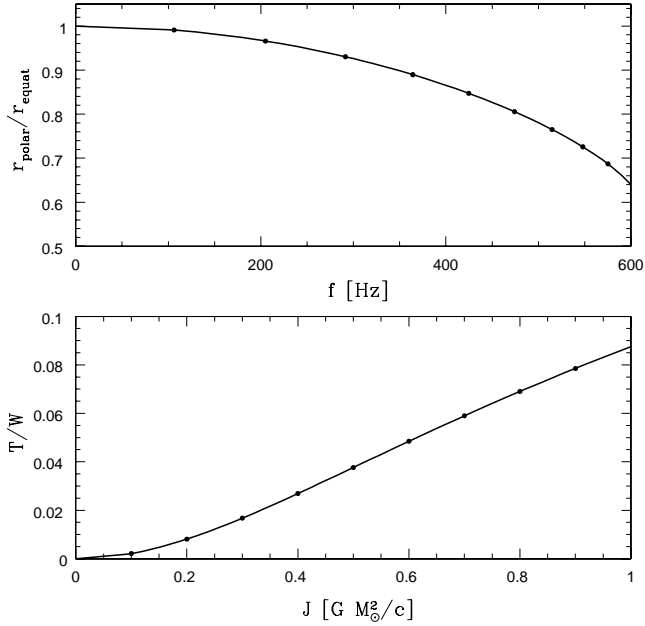}}
\caption{Top panel: the ratio of  polar radial coordinate to
the equatorial radial coordinate radius versus frequency of rotation. 
Bottom panel:
the ratio of the  kinetic energy, $T$ and the absolute value
of the potential energy, $W$ versus stellar angular momentum. 
Calculations performed for stellar configurations
consisting of the N phase of dense matter, described by the polytropic
EOS. Large dots correspond to the values of the total stellar angular
momentum, $J=(0.1,\ldots,0.9)\times GM^2_\odot/c$,
 which were used in Figs.\ \ref{fig:E.rs.poly},
 \ref{fig:E.pc.poly}.}
\label{fig:C0.obl.TW}
\end{figure}

 It should be stressed that the  configurations
${\cal C}(A,J)$ and $ {\cal C}^\star(A,J)$, considered in this
section, are  really fast   rotating ones,  not so far from
 the  Keplerian limit. This is visualized in
 Fig.\ \ref{fig:C0.obl.TW}, where  we plotted the oblateness of the
 star and the kinetic to potential energy ratio.
  And still, in spite of
 fast rotation and large oblateness, the energy release
 is the same as in a non-rotating star of the same
 initial central overpressure.
%
\subsection{Energy release  for realistic EOS}
\label{sect:E.SLy}
\begin{figure}[h]
\centering
\resizebox{3.25in}{!}{\includegraphics{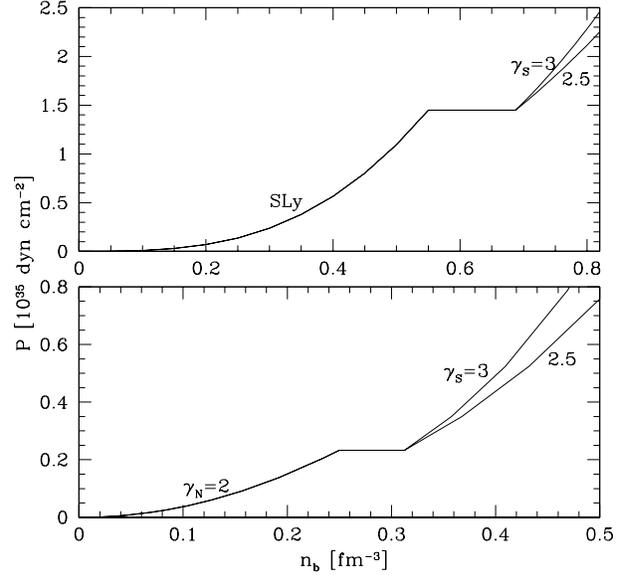}}
\caption{The EOSs with first order phase transitions, used
in the present paper.
Upper panel: the SLy EOS for the N phase, with a phase transition
to the dense S phase. We assume $n_{\rm _N}=0.54~{\rm fm^{-3}}$ and
 $\lambda_n=n_{\rm _S}/n_{\rm _N}=1.25$, which results in
$\lambda_\rho=\rho_{\rm _S}/\rho_{\rm _N}=1.28987$. For the
further description of the EOS, see the text.
Lower panel: polytropic models of the EOS with a first order
phase transition.
 We assume $n_{\rm _N}=0.25~{\rm fm^{-3}}$ and
 $\lambda_n=n_{\rm _S}/n_{\rm _N}=1.25$, which results in
$\lambda_\rho=\rho_{\rm _S}/\rho_{\rm _N}=1.2647$. For the
further description of these EOSs, see the text.
}
\label{fig:EOS.SLy.poly}
\end{figure}

%
In the present section we consider realistic EOS of the N phase.
In order to explore how a realistic rotating  neutron star will
respond to the appearance of a core of S phase core, we used
a recent  SLy EOS of Douchin \& Haensel (2001). The SLy EOS describes in
unified way (i.e., starting from a single effective nuclear
Hamiltonian) both the crust and the core of neutron star. In its
original version, the SLy EOS  assumes that neutron star core is
composed of neutrons, protons, electrons and muons. We introduced
a  softening by a first order phase transition at
$n_{\rm N}$, as shown in Fig.\ \ref{fig:EOS.SLy.poly}.

We considered two EOSs of  the S phase. Both were of  polytropic form
$P(n_{\rm b})=K_{\rm _S}n_{\rm b}^{\gamma_{\rm _S}}$,
with $\gamma_{\rm _S}=2.5$
and  $\gamma_{\rm _S}=3$, respectively. We assume the values of $n_{\rm _N}$ and $n_{\rm _S}=
\lambda_n n_{\rm _N}$. The values of $n_{\rm _N}$ and $\lambda_n$
determine then $P_0$. Equality of baryon chemical potentials
at $P=P_0$ fixes  the  other constant of the S-phase EOS,
which is the value of energy per baryon (including rest
energy) at zero pressure. In this way, the values
of $n_{\rm _N}$ and $\lambda_n$, together with $\gamma_{\rm
_S}$, fully determine the EOS of the S phase.

\begin{figure}[h]
\centering \resizebox{\hsize}{!}{\includegraphics[angle=0]{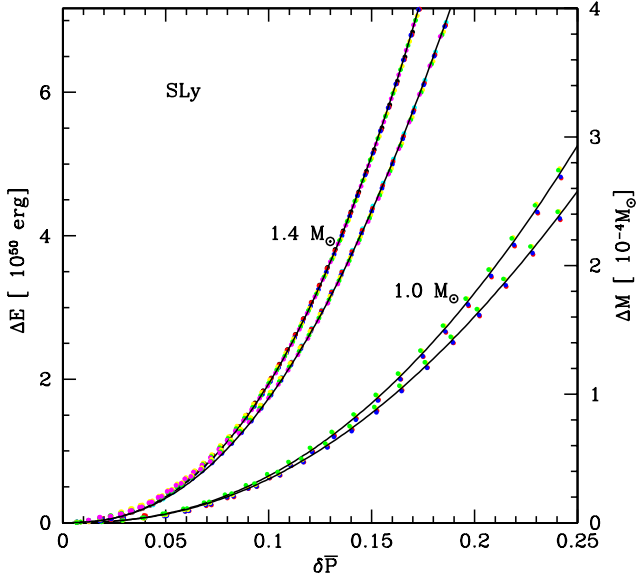}}
\caption{(Color online) The energy release due to the mini-collapse of
rotating neutron star as a function of the metastability of the normal phase of the matter
in the center of the star for the SLy EOS with first order phase transition
 located at two different points (pressures $P_0$)
for which the mass of the reference nonrotating configuration is equal
to $1.0,~1.4\msol$. For each value of $P_0$
two models of the dense core EOS are presented - polytropes with $\gamma=2.5$ (upper curve)
and $\gamma=3$. The points of different color correspond to the different values of
total angular momentum of rotating star. The results for all rotating configurations
can be very well
approximated by one curve.}
\label{fig:E.SLy}
\end{figure}

In Fig.\ \ref{fig:E.SLy} we show the energy release due to the
${\cal C}(A,J)\longrightarrow {\cal C}^\star(A,J)$ transition,
versus overpressure, for two selected values of $\gamma_{\rm _S}$. The
values obtained for different values of $J$ are marked with
different color and symbols, and can be distinguished on the
electronic versions of this paper. However, as in
the case of the polytropic models of the N phase, all color points
lie along the same line. For  a given overpressure $\delta\overline{P}$,
the energy release  does not depend on $J$ of  collapsing metastable
configuration, confirming in this way results obtained for the polytropic
EOSs. This property is  fulfilled astonishingly well for a broad range of
of stellar angular momentum, $J=(0.1,\ldots,0.7)\times G M^2_\odot/c$.

Our numerical results for different equations of state and models of a phase transitions 
presented in Figs \ref{fig:E.pc.poly}, \ref{fig:E.SLy} show that the energy release 
associated with a phase transition in the core remarkably does not depend on the stellar
rotation rate for a fixed overpressure at which the phase transition takes place.
To estimate the 'accuracy' of this conclusion we can calculate the departures of the energy
release for the rotating stars from the curve determined for nonrotating configurations.
The relative differences are of the order of 1-2\% for $\delta\overline{P}>0.1$ (except for
the smallest masses in Figs \ref{fig:E.pc.poly}, \ref{fig:E.SLy}  ($M=0.7\msol$ and $M=1\msol$ respectively)
where they can be as large as 5\%). For smaller $\delta\overline{P}>0.1$ the relative differences
are larger because the absolute value of $\Delta E$ is very small and the differences are comparable to
the numerical accuracy.


%
%

\section{Calculation of the energy release using linear response theory}
\label{sect:E.lin.resp}

Below $\rho_{\rm _N}$, the EOS for two cases (with and without phase transition)
is the same. The ambiguity of definition of ${\cal C}_0$
for rotating stars was discussed in Sect.\ 3.
We can choose a reference configuration  star with the same $J$ or the same $f$
as that of collapsing metastable one.  In fact,
 to reach the metastable configuration
with $P_{\rm c} > P_{\rm _N}$,  we have to cross the "reference
line" at the point defined by the evolutionary process leading to the instability.

We assume,  that transition of the star from a  one-phase configuration to the
configuration with a small dense core built of S phase, takes place at fixed baryon
number $A$ (no matter ejection)  and fixed total angular momentum of the star $J$
(radiation loss of $J$ neglected).
Thus for a fixed baryon number $A$ and fixed $J$,
we will calculate the  mass-energy difference between
${\cal C}$ and ${\cal C}^\star$.

We assume that the mass of the  S-phase core is much smaller
than the total stellar mass. Therefore,
from now on we will be able to restrict ourselves to a linear response of
neutron star to the appearance of the new dense S-phase core. The
calculation  is based on expressing the change in the density
profile, due to the presence to a small core, as the combination
of two independent solutions of the linearly perturbed equations
of stellar structure (Haensel et al. 1986, Zdunik et al. 1987).
The presence of a denser phase in the core changes the boundary
condition at the phase transition pressure $P_{\rm 0}$
and allows us to determine the numerical coefficients in the
expression for the density profile change. The leading term in the
perturbation of the boundary condition at the edge of the new
phase results from the mass excess due to the
higher density of the new S phase as compared to the
N one in the supercompressed core.

\subsection{Non-rotating neutron stars}
\label{sect:E.56.stat}
Let us remind expressions obtained for static case
considered by  Haensel et al. (1986), and Zdunik et al.
(1987). For a non-rotating configuration of hydrostatic
equilibrium, the density jump $\rho_{\rm _N} \to
\rho_{\rm _S}$ leads to the lowest-order expression for the
core-mass excess (with respect to the pure N-phase configuration),
\begin{equation}
\label{deltamfo}
\delta m_{\rm core} = {4\over3}\pi (\rho_{\rm _S}-\rho_{\rm _N}) r_{\rm _S}^3
 + \mathcal{O}(r_{\rm _S}^{5})~.
\end{equation}

The radius of the overcompressed N-phase core,
$r_{\rm _N}$,  is connected with the overpressure
 which can be achieved by the
N phase of matter, by
\begin{equation}
\label{eq:Pc.rN}
\delta\overline{P}=
{2\pi \over 3}G x_{\rm _N}\;\rho_{\rm _N}
\;(1+x_{\rm _N})(1+3x_{\rm _N})
r_{\rm _N}^{2}~.
\end{equation}
The conservation of the stellar baryon number during the
 ${\cal C}\longrightarrow {\cal C}^\star$ transition, implies
 a relation between $r_{\rm _N}$ and $r_{\rm _S}$ which
 neglecting terms $r_{\rm _N}^4$ and higher, reads
\begin{equation}
\label{eq:rN.rS}
(1+3x_{\rm _N})r_{\rm _N}^{2}=
(3-2\lambda_\rho +3x_{\rm _N})r_{\rm _S}^{2}-
(1-\lambda_\rho)\;a_{\rm _N}\;r_{\rm _S}^{3}
\end{equation}
where $a_{\rm _N}$ is a numerical coefficient depending on
${\cal C}_0$.

Using then Eqs.\ (\ref{eq:Pc.rN}) and (\ref{eq:rN.rS}),
one  concludes that the  leading term for the energy release
expression $\propto r_{\rm _S}^5$
contains a prefactor
$(\lambda_\rho-1)(3-2\lambda_\rho +3x_{\rm _N})$,
while the next order term $\propto r_{\rm _S}^6$ contains a
prefactor $(\lambda_\rho-1)^2$.   The
final expression for a normalized energy release reads
\begin{eqnarray}
\label{deltaQrot}
&~&\Delta\overline{E} \equiv \frac{M-M^*}{M_0}
\simeq (\lambda_\rho-1)(3-2\lambda_\rho  + 3x_{\rm _N})
\alpha_1\; {\overline{r}_{\rm _S}}^5+\cr\cr
&~&+(\lambda_\rho-1)^2\;\alpha_2\; {\overline{r}_{\rm
_S}}^6~,
\end{eqnarray}
where $\overline{r}_{\rm _S}\equiv r_{\rm _S}/R_0$.
The coefficients $\alpha_1$ and $\alpha_2$
are  functionals of the static reference configuration,
${\cal C}_0(J=0)$. Expressing now $\overline{r}_{\rm _S}$ in
terms of overpressure, we get the energy release in terms of
the overpressure,

  \begin{eqnarray}
&~&\Delta E_{50} \simeq {(\lambda_\rho -1)\over(3-2\lambda_\rho
 +3x_{\rm _N})^{1.5}}\beta_1\;(\delta\overline{P})^{2.5}
~,
  \label{eq:E.56.P}
  \end{eqnarray}
where we use standard notation $E_{50}\equiv E/10^{50}~{\rm
erg}$. The coefficient $\beta_1$
depends only
on the EOS of the N phase.  More precisely, this coefficient
 is a functional of the  {\it nonrotating} reference configuration
${\cal C}_0$ and depends rather weakly on the
mass $M_0$ of the reference configuration ${\cal C}_0$ for
a rather broad range of masses,
between $1 M_{\odot}$ and $0.8 M_{\rm max}$.
For EOSs considered in the present paper,
numerical calculations give in a good approximation
$\beta_1\simeq 0.016$.
\subsection{Rotating neutron stars}
\label{sect:E.56.rot}
Numerical results of Sect.\ 3 clearly
show that for for a given  central
overpressure of collapsing configuration, $\delta\overline{P}$,
 the energy release
does dot depend on $J$ and coincides with that for non-rotating
stars. Therefore, the formula (\ref{eq:E.56.P}) of
Sect.\ \ref{sect:E.56.stat} is valid also in the rotating
case, provided $\delta\overline{P}$ is sufficiently small.
Our calculations show that this formula is quite precise for
$\delta\overline{P}\la 0.05$~.
 The prefactor in front of $\beta_1$,  involving
$\lambda_\rho$ and  $x_{\rm _N}$,  is identical with those
obtained for  the $J=0$ transition in (Haensel et al. 1986, Zdunik et
al. 1987). The coefficient $\beta_1$  depends only
on the EOS of the N phase.  More precisely, this coefficient
 is a  functional of the  {\it nonrotating} reference configuration
${\cal C}_0$.
\section{Numerical estimates of the energy release}
\label{sect:num.estimates}
For practical application, and for small overpressures,
$\delta\overline{P}< 0.1$,  it is convenient to
summarize results obtained in  Sects.\ \ref{sect:E.poly}-\ref{sect:E.SLy}
in a formula
\begin{equation}
\Delta E_{50}= a_1\left(\delta\overline{P}\right)^{2.5}
\label{eq:General.num}
\end{equation}
The only dependence of coefficient $a_1$ on the
phase transition parameter $\lambda_\rho$ is via
prefactors,
\begin{equation}
a_1={(\lambda_\rho-1)\over(3-2\lambda_\rho+3x_{\rm _N})^{1.5}}\beta_1~,
\label{eq:a1.a2.scaling}
\end{equation}
which allows for a rapid re-calculation of $a_1$  when one changes
the value of the density jump.

For the phase transition model considered in Sects.\
\ref{sect:E.SLy}, the energy accompanying
phase transition in a metastable star with central overpressure
$\delta\overline{P}=0.1$ is about $2\times 10^{50}~$erg, and
becomes one order of magnitude smaller for
$\delta\overline{P}=0.05$

%
\section{Discussion and conclusions}
\label{sect:conclusions}
%
The most important result of the present paper is that the
total energy release associated with a first order phase
transition at the center of a rotating neutron star does
depend only on the overpressure at the center of the
metastable configuration  and is {\it independent of the star
rotation rate}.  This result holds even for fast stellar rotation,
when the star shape deviates significantly from sphericity, and
$J\simeq 0.9GM_\odot^2/c$, and for overpressures as high as (10-20)\%.
 The property is of great practical
importance, because it implies that the calculation of the
energy release for a given overpressure, requiring very high precision
to guarantee $A=A^\star$,  can be reduced to the case of
non-rotating spherically symmetric stars.

To illustrate the practical importance of our result, let us
consider the case of overpressure $\delta\overline{P}=0.03$,
when the energy release is $2\times 10^{49}~$erg. For a
$1.4~M_\odot$ star this constitutes about $10^{-5}$ of the
stellar mass-energy. Therefore, to get a meaningful result for the
energy  of this magnitude, conservation of the total baryon
number has to be satisfied at the level of one part in  a
million, which is next  to impossible for a 2-D
calculation with a realistic EOS. However, such a precision
can be easily reached for spherical stars, provided the EOS is
used in a thermodynamically consistent way (see, e.g., Haensel
\& Proszynski 1982).

Moreover, we have shown that for overpressures smaller than
5\%, the energy release is proportional to
power $2.5$ of the overpressure, with coefficient
weakly dependent on the mass of collapsing configuration, for
stellar masses  in
the range from $1\;M_{\odot}$ to about $0.8\;M_{\rm max}$.

The energy release $\Delta E$  that we calculated is an
absolute upper bound on the energies which can released
 as a result of a phase transition at the star
center. The available channels may include, for example, stellar
pulsations, gravitational radiation, heating of stellar
interior, and X-ray emission from neutron star surface.
Moreover, the phase transition in rotating star
implies shrinking of stellar radius, decrease of moment of
inertia, and spin-up of rotation. These topics will
be considered in  our next paper.
%
\acknowledgements{ This work was partially supported by the
Polish MEiN grant no. 1P03D.008.27 and by the LEA
Astrophysics Poland-France
(Astro-PF) program. MB was partially supported by the Marie Curie
Intra-european Fellowship MEIF-CT-2005-023644
within the 6th European Community Framework
Programme.}



\begin{thebibliography}{}
\bibitem[2005]{bejger2005}
Bejger M., Haensel P., Zdunik J.L., 2005, MNRAS, 359, 699
\bibitem[2001]{sly}
Douchin F., Haensel P., 2001, A\&A, 380, 151
\bibitem[1997]{Glend.book}
Glendenning N. K., 2000, Compact Stars. Nuclear Physics, Particle
Physics, and General Relativity, Springer, Berlin
\bibitem[1982]{HaenProsz82}
Haensel P., Proszynski, M., 1982, ApJ, 258, 306
\bibitem[1986]{HZS}
Haensel P., Zdunik J. L., Schaeffer R., 1986, A\&A, 160, 251
\bibitem[1997]{IidaSato1997}
Iida K., Sato K., 1997,  Prog. Theor. Phys., 98, 277
\bibitem[1981]{Kaempfer1981}
Kaempfer, B., 1981, Phys. Lett. 101B, 366
\bibitem[1983]{SHZ1983}
Schaeffer R., Haensel P., Zdunik J.L., 1983, A\&A, 126, 121
\bibitem[1971]{Seidov1971}
Seidov Z.F., 1971, Sov. Astron.- Astron.Zh., 15, 347
\bibitem[1999]{weber.book}
Weber F., 1999, Pulsars as Astrophysical Laboratories for Nuclear
and Particle Physics, IoP Publishing, Bristol \& Philadelphia
\bibitem[Zdunik et al. (1987)]{ZHS}
Zdunik J. L., Haensel P., Schaeffer R., 1987, A\&A, 172, 95
\bibitem[Zdunik et al. (2006)]{ZBHG}
Zdunik J. L., Bejger M., Haensel P., Gourgoulhon E., 2006, A\&A, 450, 747
\end{thebibliography}
\end{document}